\documentclass[twocolumn,pre,showpacs,superscriptaddress]{revtex4}

\usepackage{graphicx}
\usepackage{amsmath}

\begin{document}

\title{One dimensional Brownian motion in hard rods: adiabatic piston problem }

\author{ M. Ebrahim Foulaadvand$^{1,2}$ and  M. Mehdi Shafiee $^{1}$ \\
foolad@iasbs.ac.ir\\
{\small $^1$ Department of Physics, University of Zanjan, P. O. Box 313, Zanjan, Iran } \\
{\small $^2$ School of Nano-science, Institute for Research in Fundamental Sciences (IPM) , P.O. Box 19395-5531, Teheran, Iran} \\
 }
\date{\today}

\begin{abstract}

We have investigated the motion characteristics of a movable piston immersed in a one dimensional gas of hard rods by event-oriented molecular dynamics in the absence of thermal noise. Periodic and reflecting boundary conditions are explored. It is shown that the piston undergoes systematic oscillations with decaying amplitudes in short times before it comes to global thermodynamic equilibrium. Moreover, the diffusion of the piston is explored and analytical expressions for its equilibrium mean-squared displacement is obtained. It is shown that MSD of the piston does not differ much from the normal rods despite its mass and length are significantly larger.

\end{abstract}
\maketitle

%\keywords{ Hard rods, adiabatic piston, diffusion, event oriented MD, density wave }

\section{introduction}

The rather old but controversial problem of {\it adiabatic piston} \cite{calen,lebowitz,alkemade} in equilibrium thermodynamics has recently given a renewed interest to some statistical physicist especially after the works of Lieb \cite{lieb1,lieb2}. The problem consists of an isolated cylinder with two compartments, separated by an insulating piston which is free to move along the cylinder axis of symmetry \cite{gruber1,piasecki1,piasecki2,frachebourgh,kestemont,munakata,chernov,white,taniguchi,gruber2,gruber3,brito,redner}. Feynman \cite{feynman} discusses this example and gives several hand-waving arguments to convince the reader that the piston will indeed performs a directed motion under certain conditions. Piaseckia and Gruber proposed a simple one dimensional model to mimic the basic features of the problem \cite{piasecki1,piasecki2}. A massive movable piston with mass $M$ separates left and right segments ideal gases and is subjected to elastic collisions with gases molecules. Solving the linearised Boltzmann equation, it was interestingly shown that if the temperatures of the left and right gases differ, despite having the same pressure $P$, the piston acquires a net non zero velocity towards the warmer segment. In the case where the temperature of the fluids on both sides of the piston are equal the problem reduces to the classical Brownian motion of the {\it Rayleigh piston}, which has been extensively studied \cite{hoare,driessler}. {\bf Gruber and Morris gave a detailed analysis of the 2D version of the problem under the influence of an external force on the piston} \cite{gruber3}. Kestemont et al presented a two dimensional model in which the fluid particles were hard disks (Enskog gas) that collided elastically with the piston (a vertical line in 2D) \cite{kestemont}. Their simulations revealed some novel aspects the most important of which was the damped oscillatory motion of the piston \cite{kestemont,frachebourgh}. More recently the two dimensional piston problem was re examined by White et al \cite{white}. By time series and spectral analysis of the piston position, they managed to compute the frequency as well as the damping constant of the piston oscillations. {\bf The examination of damping coefficient and oscillation frequency were nicely done by Malek Mansour et al in a hydrodynamic description} \cite{mansour1}. Besides piston problem, recently a new stride has been opened in view of the empirical importance of the subject of transport in quasi one dimensional channels \cite{lizana1,barkai,fouladvand}. In this paper we consider a one dimensional piston problem. The main difference of our model to the preceding ones is that our fluid particles are not point-like but are rods. This problem was originally introduced by Tonks \cite{tonks} and the fluid is known as {\it Tonks gas} in the literature. In this paper we focus our attention to the motion and diffusion properties of a tracer particle (the piston) under a fully deterministic hard core potential among rods and the piston. The tracer mass and length notably differ from other normal rods. The piston motion mimics the motion of a Brownian particle immersed in a gas of smaller rods. From the theoretical perspective, we also hope that our investigations shed more light on collective phenomena that arise in 1D fluids.

\section{Description of the problem}

Imagine $N$ one dimensional rods each having a length $l$ and a mass $m$. They are restricted to move along a straight line. The rods interact via a hard core impulsive potential which implies that each rod moves with a constant velocity between elastic collisions. We consider two types of boundary conditions: periodic and reflecting. Moreover, the collisions are assumed to be elastic. See figure (1) for illustration. Recall that in an elastic collision between two identical particles in 1D they exchange their velocities. Since the total energy is purely kinetic, the temperature remains constant as given in the initial condition. Lengths and masses are scaled in rod length $l$ and in rod mass $m$ which are taken as unity throughout the paper. Time is measured in thermal unit i.e.; $\tau=l\sqrt{\frac{m}{k_BT}}$. We take the system length $L$ and denote the number density by $\rho=\frac{N}{L}$. The dimensionless packing fraction $\eta$ is related to number density as $\eta=l\rho$ and is restricted between zero and one.

\begin{figure}[h]
\centering
\includegraphics[width=7.5cm,height=1.5cm,angle=0]{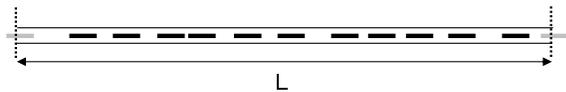}
\caption{A one dimensional gas of hard rods interacting via a
repulsive hard-core potential. Collisions among rods are elastic and periodic boundary condition is imposed.}\label{Fig1}
\end{figure}

We now consider the effect of inserting a piston in the system. By piston we mean a rod with notably larger size and mass. The piston or the Brownian particle is subjected to incessant collisions by normal rods and performs a seemingly stochastic motion. Our piston, placed initially in the middle of the system, is a rod with a length $l_p$ and mass $M$ which are $s$ times larger than a normal rod: $l_p=sl$ and $M=sm~ (s>1$). The post-collision velocities ($V'$ and $v'$) after a normal rod elastically collides the piston are simply read from pre-collision ones $v$ and $V$ via conservation laws of energy and momentum:

\begin{eqnarray}
V'=V-\frac{2m}{m+M}(V-v)=V-\frac{2}{1+s}(V-v)
\end{eqnarray}

\begin{eqnarray}
v'=v+\frac{2M}{m+M}(V-v)=v+\frac{2s}{1+s}(V-v)
\end{eqnarray}

After experiencing sufficient collisions with normal rods, the piston thermalises with the gases on its left and right sides an acquires the system temperature $kT$. Before simulating this problem, it would be instructive to analytically compute the mean-squared displacements (MSD) $\langle [\Delta x(t)]^2 \rangle=\langle [x(t)-x(0)]^2\rangle$. We shall do it for the reflecting boundary condition. Suppose our $N$ rods are restricted to move within a one dimensional line of length $L$ with reflecting boundaries. The origin $x=0$ is taken at the left end. The canonical partition function can be written as follows:

$$Z_N(L,T)=\int_{\frac{l}{2}}^{L-Nl+\frac{l}{2}} dx_1
\int_{x_1+l}^{x_3-l} dx_2 \times \cdots \times \int_{x_{N-1} +l}^{L-\frac{l}{2}} dx_N $$

Straightforward integration gives the partition function as $Z_N(L,T)=\frac{1}{(N)!}(L-Nl)^N$.
Comparison to partition function of the periodic boundary condition  $Z_N^{c}(L,T)=\frac{L(L-Nl)^{N-1}}{(N-1)!}$ reveals the difference induced by the type of boundary condition \cite{bishop}. The averages however, should remain unchanged in the thermodynamics limit. Having evaluated $Z$ we are now able to compute the mean-squared displacement of any particle. We only express the result and present the details elsewhere. For the $m$-th particle we have:
$\langle x_m \rangle=(m-\frac{1}{2})l +\frac{m\Phi}{N+1}$. This result is in agreement with our intuition.
In fact $g=\frac{\Phi}{N+1}$ is the average gap between rods when they are placed equi distantly relative to each other and the walls. It simply states that the average position of rods coincides with the equi distance configuration. For the mean-squared $\langle x_m^2 \rangle$ we find: $\langle x_m^2 \rangle=(m-\frac{1}{2})^2l^2 + \frac{2m(m-\frac{1}{2})l\Phi }{N+1} + \frac{(m^2+m)\Phi^2}{(N+1)(N+2)}$. The final stage will be evaluating $\langle (\Delta x_m)^2 \rangle=\langle x_m^2 \rangle -\langle x_m \rangle^2$. It turns out:

\begin{eqnarray}
\langle (\Delta x_m)^2
\rangle=\frac{2mN^2(N-m+1)l^2}{(N+1)^2(N+2)}(\frac{1-\eta}{\eta})^2
\end{eqnarray}

Apparently the saturation value of MSD depends on number of rods. This is in contrast to higher dimensions where MSD of particles are identical. Moreover, the symmetry $ m \Leftrightarrow N-m+1 $ is evident. Adjacent rods to the reflecting boundaries i.e.; rods $1$ and $N$ posses the smallest saturated value of MSD:$ \langle (\Delta x_1)^2 \rangle=\langle (\Delta x_N)^2 \rangle=l^2(\frac{1-\eta}{\eta})^2 $. The middle rod, $m=\frac{N+1}{2}$, has the largest saturated value of MSD:
$ \langle (\Delta x_{mid})^2 \rangle=\frac{N}{4}l^2(\frac{1-\eta}{\eta})^2 $. Figure (2) shows the analytical steady state value of MSD for different rods.

\begin{figure}[h]
\centering
\includegraphics[width=7.5cm,height=5.5cm,angle=0]{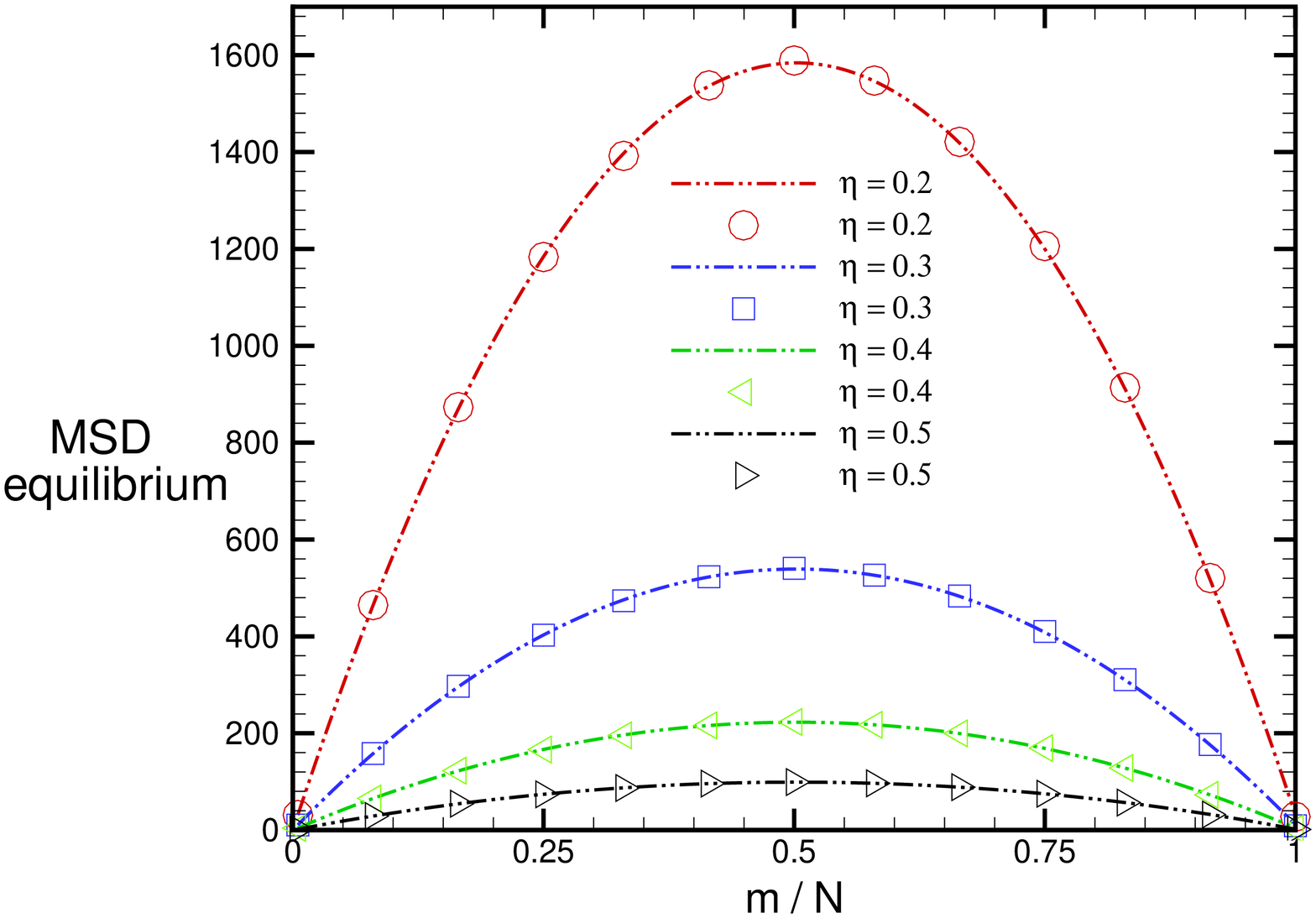}
\caption{Comparison of analytical saturation values of $\langle (\Delta x)^2 \rangle$ for all the rods with the corresponding computed ones from molecular dynamics simulations. The boundary condition is reflecting, number of rods has been $N=200$ with $N_{col}=1 \times 10^6$ per rod. Solid lines show analytical results whereas dotted lines correspond to simulation data.}\label{Fig2}
\end{figure}

\section{Simulation of the piston problem}

In this section we consider the simulation of the piston problem. The system evolves in an event-oriented molecular dynamics i.e.; the time elapses collision-by-collision \cite{bishop}. We initially place the rods in a random manner (no overlapping). The piston is initially fixed (immobile) in the system middle. Initial rods velocities are taken from a Gaussian distribution centred at zero with a standard deviation $\sigma=k_BT$. In the paper we take $k_BT=1$. We have extensively performed runs having various initial conditions. Their outcome are similar to each other within statistical errors. All the results have been obtained over a single run unless otherwise stated. Let us present our simulation data. We first consider the periodic boundary condition. Figures (3) exhibits the time evolution of the piston MSD for $s=10$ and $s=100$ each for various values of $\eta$.\\

\begin{figure}[h]
\centering
\includegraphics[width=7.5cm,height=5.5cm,angle=0]{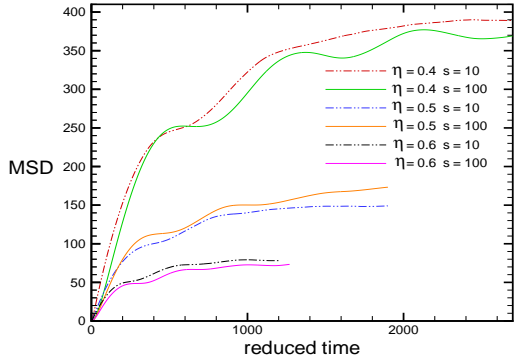}
\caption{Simulation data of $\langle (\Delta x)^2 \rangle$ of a single piston ($s=10$ and $s=100$) for
various $\eta$ and $N=1000$. The boundary condition is periodic and the piston has been set as the middle rod.}\label{Fig3}
\end{figure}

Our results demonstrate that $\langle (\Delta x)^2 \rangle$ shows a nontrivial and interesting behaviour. For short times it increases linearly in time. In the intermediate times the MSD undergoes smooth oscillations. For larger $\eta$ it becomes almost saturated. Our simulations have been executed with $N=1000$ rods and each rod has, on average, experienced one million of collisions with its neighbours. The same results obtained for two millions collisions per rod (not shown here) are almost identical to the results of one million collisions. This confirms the time oscillations are meaningful and are not due to poor statistics. Furthermore, our simulations (not shown here) shows that these temporal oscillations are not artifacts of finite size. We speculate that these fluctuations are associated to formation of standing sound waves (cavity modes) generated by the density fluctuations in the system
\cite{kestemont,white,mansour1,gutkowicz1,gutkowicz2,mansour2}. As a matter of fact, the MSD oscillation roots in the oscillatory behaviour of the piston itself. In a similar 2D problem of adiabatic piston \cite{kestemont,white}, a mobile piston separates two gases of hard disks. It has been shown that before reaching to the global thermodynamics equilibrium the piston undergoes systematic damped oscillations in short time scale with a characteristic frequency $\omega$. The density wave $\xi(x,t)$, which drives the piston, obeys the d'Alembert's wave equation $\rho_m \frac{\partial^2 \xi(x,t)}{\partial t^2}=K\frac{\partial^2 \xi(x,t)}{\partial x^2}$ in which $K=-V(\frac{\partial P}{\partial V})_s$ is the adiabatic bulk modulus and $\rho_m$ is the gas mass density. The short time oscillations of the piston has also been numerically reported in a simple three-particle toy model version of the piston problem \cite{redner}. In our problem the MSD oscillations amplitude become enhanced when the piston becomes larger. It is also noticeable that despite the larger piston ($s=100$) is one order of magnitude larger than the smaller one ($s=10$) their MSD are close to each other. We now turn to reflecting boundary condition. Suppose the piston is the $M$th rod. Calculations for the saturation value of the piston MSD are similar to the pure system. The final result yields to be: $Z=\frac{[\Phi-(s-1)l]^N}{N!} $. Straightforward but lengthy manipulations gives the saturation value of the piston MSD as follows:

\begin{eqnarray}
\langle (\Delta x_M)^2\rangle=\frac{2MN^2(N-M+1)l^2}{(N+2)(N+1)^2}(\frac{1-\eta}{\eta}-\frac{s-1}{N})^2
\end{eqnarray}

The symmetry $M \Leftrightarrow N-M+1$ is evident in (4). In the thermodynamics limit there will be no dependence on piston size $s$. Figure (4) sketches the piston MSD for various values of $s$ versus time. Analogous to the pure system the MSD becomes saturated which is due to finiteness of the system size. The systematic oscillations in short time are noticeable. These damped oscillatory motions are diminished after mechanical equilibrium is established. Larger amplitudes are associated to larger pistons. As you see the period of oscillations does not significantly depend on $s$. A similar observation has been reported in
the oscillation of piston in a 2D problem \cite{kestemont} and theoretically analysed in \cite{white,mansour1}.

\begin{figure}[h]
\centering
\includegraphics[width=7.5cm,height=5.5cm,angle=0]{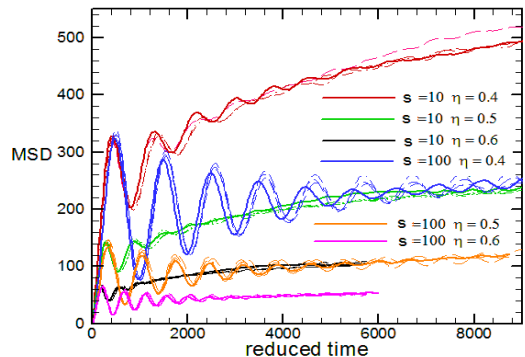}
\caption{$\langle (\Delta x)^2 \rangle$ of a single piston for
various $\eta$ with $s=10$,$s=100$ and $N=500$ in the reflecting boundary condition.
The results of three independent runs are shown together for each $s$ and $\eta$.
Analytical equation (5) gives $\langle (\Delta x_M)^2 \rangle=562,249,111$ for $\eta=0.4,0.5,0.6$
and $s=10$ respectively which are in a good agreement with simulation.}\label{Fig1}
\end{figure}

In order to convince that MSD oscillations are not artifact of finite size we have computed it for various system size at a fixed $\eta=0.4$. Figure (5) illustrates this situation and you see that amplitudes survive for larger system size. We observe the MSD oscillations do not disappear for larger $N$. We have simultaneously exhibited three runs associated to different initial velocities. The graphs confirm that oscillations are not influenced by the initial conditions.

\begin{figure}[h]
\centering
\includegraphics[width=7.5cm,height=5.5cm,angle=0]{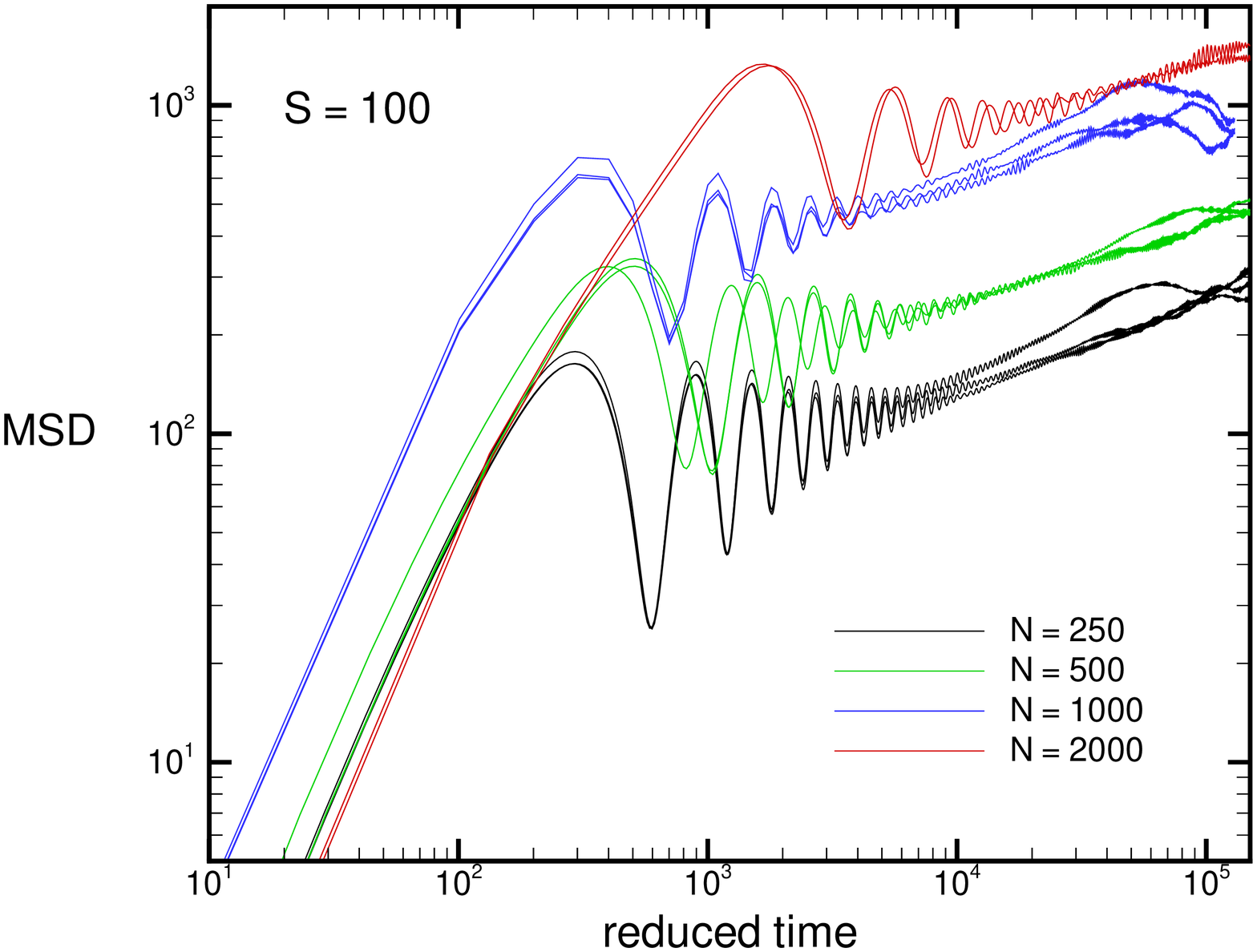}
\caption{The MSD of a $s=100$ piston for various system sizes at $\eta=0.4$ in reflecting boundary condition. For each $N$ the results of three independent runs each having different initial conditions are shown.}\label{Fig1}
\end{figure}

The type of boundary conditions affects the oscillations features substantially. In fact, the reflecting boundary condition highly amplifies the oscillations. Let us give a quantitative explanation for the piston oscillations. The piston divides the gas into two segments. Each segment resembles a Tonk gas confined between a wall and the massive piston. By fluctuations, the piston moves and this causes one of the
segments to expand whereas the other one contracts. The compressed Tonks gas exerts restoring force which tries to push the piston backward. Consequently it undergo an oscillatory motion. By the assumption that the piston motion generates a quasi standing density wave $\xi(x,t)$ in the left and right fluids, it would be plausible to assume that the piston is driven by the sound waves. Consequently one can write the following equation of motion for the piston \cite{white}:

\begin{eqnarray}
M\frac{d^2X}{d t^2}=-K\frac{\partial \xi_L}{\partial x}|_{x=X-\frac{l_p}{2}} + K\frac{\partial \xi_R}{\partial x}|_{x=X+\frac{l_p}{2}}
\end{eqnarray}

Here $\xi_L(x,t)$ and $\xi_R(x,t)$ are the generated sound waves in the left and right fluids and $X$ the piston centre of mass. We estimate the theoretical value of the oscillations frequency $\omega$ by $ck$ in which $c$ is the sound velocity and $k$ the wave number. Once the system equation of state is known one can evaluate the isothermal compressiblity $K_T$ by the formula $K_T=-V(\frac{\partial p}{\partial V})_T$. In our 1D model it reduces to:

\begin{eqnarray}
K=-L(\frac{\partial p}{\partial L})_T
\end{eqnarray}

To proceed, we need to know the equation of state which expresses the pressure $p$ in terms of system length $L$. For the Tonks gas with $N$ rods we have \cite{percus74}:

\begin{eqnarray}
p=\frac{NkT}{L-Nl}
\end{eqnarray}

After taking the derivative we find the isothermal bulk modulus $K_T$ as follows:

\begin{eqnarray}
K_T=\frac{\eta}{1-\eta}\frac{kT}{l}
\end{eqnarray}

Approximating the isentropic compressibilty $K_s$ by $K_T$ we find the sound velocity as follows:

\begin{eqnarray}
c=\sqrt{\frac{\eta}{1-\eta}\frac{kT}{\rho_m l}}=\sqrt{ \frac{kT}{m(1-\eta)} }
\end{eqnarray}

It remains to determine the wave number $k$. We estimate it by assuming that the only the lowest mode is exited. In this mode we take the piston to be fixed at its equilibrium position $\frac{L}{2}$. Therefore, in the lowest mode we have $\frac{L}{2}=\frac{\lambda}{2}$ which gives $\lambda=L$. Consequently we find: $k=\frac{2\pi}{\lambda}=\frac{2\pi}{L}$. We can now estimate the period of oscillations via $\omega=ck$. Noting that $kT=m=1$ equation (9) gives $c=\frac{1}{\sqrt{1-\eta}}$. This give the angular frequency:

\begin{eqnarray}
\omega=\frac{1}{\sqrt{1-\eta}}\frac{2\pi}{L}=\frac{2\pi\eta}{Nl\sqrt{1-\eta}}
\end{eqnarray}

Our naive estimation for the oscillations period $T$ turns out to be:

\begin{eqnarray}
T=\frac{2\pi}{\omega}=\frac{\sqrt{1-\eta}}{\eta}Nl
\end{eqnarray}

Taking $N=500$ and $\eta=0.4$ gives $T=970$ which is a qualitative agreement with simulation data of figure (4). {\bf In reference \cite{mansour1} a macroscopic equation, which includes a damping term, for the motion of the two-dimensional adiabatic piston immersed in hard disks fluid is derived within a hydrodynamics approach. This approach nicely gives the characteristics (relaxation time and oscillation period) of damped oscillatory motion of the piston towards a final equilibrium. It would be an interesting task to employ this approach in one dimension}. Another interesting point is that not only the piston but also the other normal rods undergo such oscillations at short times Fig. (6) sketches the MSD time evolution of some neighbouring rods to the piston. The oscillations are in phase to each other. This suggest the existence of a sort of collective excitements i.e.; the standing modes in the system. The piston acts as a slow moving boundary which regulates and coordinates the motion of normal rods. \\

\begin{figure}[h]
\centering
\includegraphics[width=7.5cm,height=5.5cm,angle=0]{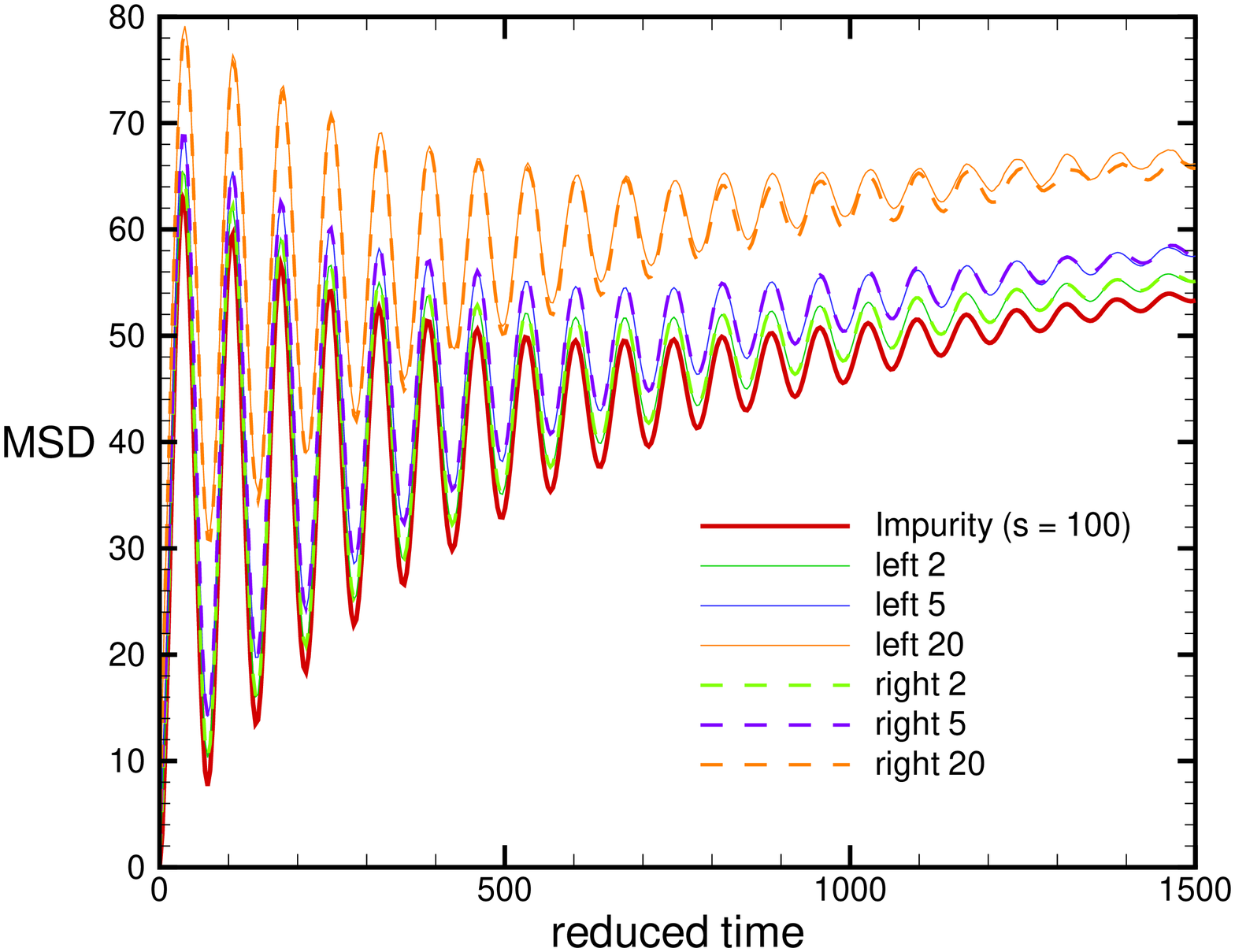}
\caption{Short time behaviour of MSD of adjacent rods to the piston at its both sides. The numbers in front of the words {\it left } and {\it right} (inside figure) refer to the rod number from the the piston .}\label{Fig1}
\end{figure}

\section{Summary and conclusion}

We have explored the diffusion characteristics of piston immersed in a one dimensional gas of hard rods by event-oriented molecular dynamics simulation. The interaction between rods is assumed to be hard core and no thermal noise exists. Two boundary conditions namely periodic and reflecting are investigated. Despite the huge difference of mass and length between the piston and normal rods, its MSD differs only slightly with the MSD of normal rods. Another notable aspect is the oscillatory behaviour of the piston in short times. Besides the piston, other normal rods will exhibit similar oscillations which are in phase and coordination with the piston. We speculate these oscillations are linked to collective excitements of the density wave. It is shown that the reflecting boundaries intensifies the MSD oscillation amplitudes. The oscillations period is theoretically obtained. Besides numerics, we have also analytically computed the saturation value of MSD in canonical ensemble theory for the reflecting boundary condition.

\section{Acknowledgement}

We are highly indebted to Dr. Reza Ejtehadi from Sharif (former Ariya Mehr) university in Tehran for enlightening and fruitful discussions.
Our gratitude is expressed to Eli Barkai and Tobias Ambj\"{o}rnsson for useful comments and discussions.
We thank Alireza Saffar Zadeh for useful helps. We wish to express our gratitude to anonymous referees for their valuable comments and suggestions.
%Fruitful comments of an anonymous referee of {\it Journal of statistical mechanics} is appreciated.

\end{document}